\documentclass[aps,prc,preprint,groupedaddress]{revtex4-1}

\usepackage{color}
\usepackage{graphicx}
\usepackage{hyperref}
\usepackage{lineno}

\begin{document}

\title{\boldmath 
Predictions for nuclear structure functions from the impact-parameter dependent Balitsky-Kovchegov equation
\unboldmath}


\author{J. Cepila}
\author{J. G. Contreras}
\author{M. Matas}
\affiliation{Faculty of Nuclear Sciences and Physical Engineering,
Czech Technical University in Prague, Czech Republic}
\date{\today}

\begin{abstract}
In this work we present dipole scattering amplitudes, including the dependence on the impact-parameter, for a variety of nuclear targets of interest for the electron-ion colliders (EICs) being currently designed. These amplitudes are obtained by numerically solving the Balitsky-Kovchegov equation with the collinearly improved kernel. Two different cases are studied: initial conditions representing the nucleus under consideration and the solutions based on an initial condition representing a proton complemented by a Glauber-Gribov prescription to obtain dipole-nucleus amplitudes. We find that the energy evolution of these two approaches differ. We use the obtained dipole scattering amplitudes to predict  ($i$) nuclear structure functions that can be measured in deep-inelastic scattering at EICs and ($ii$) nuclear suppression factors that reveal the energy evolution of shadowing for the different cases we studied. We compare our predictions with the available data.
\end{abstract}


\maketitle


\section{Introduction}
Feasibility studies for  electron-ion colliders (EICs), like those proposed in the USA~\cite{Accardi:2012qut} or at CERN~\cite{AbelleiraFernandez:2012cc}, are an essential ingredient towards the design of these machines. Inclusive measurements are among the most important observables in this context. In particular, the study of the structure function $F^A_2(x,Q^2)$ at small Bjorken-$x$ for photons of virtuality $Q^2$ at a pertubative scale, and for a variety of nuclei $A$, is expected to yield a new understanding of the high-energy limit of quantum chromodynamics (QCD). Comparison of these measurements with those reported by  H1 and ZEUS~\cite{Abramowicz:2015mha}  for the corresponding structure function of the proton, $F^p_2(x,Q^2)$, promise to shed new light on the origin of shadowing, the phenomenon that the parton distributions of nucleons bounded in a nucleus are suppressed with respect to those of  free nucleons~\cite{Armesto:2006ph}.

At small values of $x$, the dominant parton distribution is that of gluons; thus the case of gluon shadowing has been the focus of attention for theorists since a long time; e.g.,~\cite{Mueller:1985wy,Nikolaev:1990ja}. A  process expected to occur in this kinematic regime is saturation, namely the fact that the density of gluons is so high that they start to interact with each other, even in the domain of perturbative QCD. (For a recent review see~\cite{Albacete:2014fwa}.) An early equation to describe saturation was introduced in the seminal work~\cite{Gribov:1984tu}, while nowadays it is common to use the Balitsky-Kovchegov (BK) equation for this type of studies. The leading order BK equation, discussed in Sec~\ref{sec:Formalism}, was derived  in~\cite{Balitsky:1995ub} and~\cite{Kovchegov:1999yj} using two  independent approaches. Later on, corrections to account for the running of the coupling~\cite{Kovchegov:2006vj,Albacete:2007yr} as well as the resummation of other logarithmic contributions~\cite{Iancu:2015joa,Iancu:2015vea} were incorporated into this formalism. In the approximation of considering a large homogeneous target, that is, disregarding the impact parameter dependence, this equation has been successfully used to describe the existing $F^{p}_2(x,Q^2)$ data, e.g. in~\cite{Albacete:2010sy,Iancu:2015vea}.

The first attempt at solving the BK equation including the impact-parameter dependence~\cite{GolecBiernat:2003ym} found that the solutions developed so-called Coulomb tails: an unphysical grow of the amplitude at large impact parameters. Nonetheless, using some extra ad hoc corrections it was possible to describe the structure function data of the proton~\cite{Berger:2010sh,Berger:2011ew}. Recently, our group discovered that using the collinearly-improved kernel introduced in~\cite{Iancu:2015vea} the problem of Coulomb tails is tamed such that a successful phenomenology using the BK equation is possible~\cite{Cepila:2018faq,Bendova:2019psy}.

In this article, we solve  the BK equation with the collinearly improved kernel for  different nuclei of importance for future EICs and predict their structure functions as well as the corresponding nuclear suppression factors, which are a direct measurement of shadowing. We study two cases: solutions obtained from an initial condition representing the nuclei (denoted as b-BK-A below), and solutions of the proton case coupled to a Glauber-Gribov prescription to obtain the nuclear structure functions (denoted as b-BK-GG). Other approaches to the computation of nuclear structure functions can be found for example in~\cite{Armesto:2002ny,Cazaroto:2008iy,Agozzino:2014qta,Marquet:2017bga,Aschenauer:2017oxs}

The rest of the text is organised as follows: Sec.~\ref{sec:Formalism} contains a brief review of the formalism, including the definition of the initial conditions and the values of the parameters used in the computation. Section~\ref{sec:DipoleAmplitude} discusses the behaviour of the dipole scattering amplitudes obtained by solving the BK equation for the different nuclei and the b-BK-A and b-BK-GG approaches. Section~\ref{sec:Predictions} presents our predictions for the nuclear structure functions and nuclear suppression factors for all cases under study as well as a comparison with the available data. Finally, in Sec.~\ref{sec:Summary} we provide a brief summary of the presented work as well as an outlook of future steps.

\section{Formalism
\label{sec:Formalism}}
\subsection{The Balistsky-Kovchegov equation with the collinearly improved kernel}
The leading order Balitsky-Kovchegov equation~\cite{Balitsky:1995ub,Kovchegov:1999yj} is
\begin{eqnarray}\label{fullbalitsky}
\frac{\partial N(\vec{r}, \vec{b}, Y)}{\partial Y} = \int d\vec{r_{1}}K(r,r_{1},r_{2})& \Big(& N(\vec{r_{1}}, \vec{b_1}, Y) + N(\vec{r_{2}}, \vec{b_2}, Y) - N(\vec{r}, \vec{b}, Y) \nonumber \\
& & - N(\vec{r_{1}}, \vec{b_1}, Y)N(\vec{r_{2}}, \vec{b_2}, Y)\Big).
\end{eqnarray}
It describes the  evolution in rapidity $Y$ of the dipole scattering amplitude $N(\vec{r}, \vec{b}, Y)$. Here, the sizes of the mother and daughter dipoles are 
 $r\equiv|\vec{r}\,|$, $r_1\equiv|\vec{r_1}|$, and $r_2\equiv|\vec{r_2}|\equiv|\vec{r}-\vec{r_1}|$, respectively. The  magnitudes of the corresponding impact parameters between these dipoles and the hadronic target are $b\equiv|\vec{b}|$, $b_1\equiv|\vec{b_1}|$, $b_2\equiv|\vec{b_2}|$. All these vectors are two-dimensional and live in the impact-parameter plane.
 
 We  solve the equation for the case when the following two conditions are fulfilled, ($i$) the evolution depends only on the magnitude of both the dipole size and the impact-parameter vectors, and  ($ii$)  the angle between $\vec{r}$ and $\vec{b}$ is fixed to zero: 
 \begin{eqnarray}
\frac{\partial N(r,b, Y)}{\partial Y} = \int d\vec{r_{1}}K(r,r_{1},r_{2}) & \Big( & N(r_1,b_1, Y) + N(r_2,b_2, Y) - N(r,b, Y) \nonumber \\
& & - N(r_1,b_1, Y)N(r_2,b_2, Y) \Big).
\label{BKused}
\end{eqnarray}

For the kernel we use the recently proposed collinearly improved version~\cite{Iancu:2015joa}
\begin{equation}\label{collinearlyimproved}
K_{\rm ci}(r, r_1, r_2) = \frac{\overline{\alpha}_s}{2\pi}\frac{r^{2}}{r_{1}^{2}r_{2}^{2}} \left[\frac{r^{2}}{\min(r_{1}^{2}, r_{2}^{2})}\right]^{\pm \overline{\alpha}_sA_1} K_{\rm DLA}(\sqrt{L_{r_1r}L_{r_2r}}),
\end{equation}
where (see also~\cite{Vera:2005jt})
\begin{equation}
K_{\rm DLA}(\rho) = \frac{J_1(2\sqrt{\overline{\alpha}_s \rho^2})}{\sqrt{\overline{\alpha}_s \rho}},
\label{eq:DLA}
\end{equation}
$J_1$ is the Bessel function, the anomalous dimension is $A_1= 11/12$, and $
L_{r_ir} = \ln\left(r_i^2/r^2\right)$.
The sign is positive when the size of the original dipole is smaller than the size of each of the daughter dipoles
and  negative  otherwise. The smallest dipole prescription is used for the running coupling: $\overline{\alpha}_s = \alpha_s (r_{\min})N_c/\pi$, where $r_{\min} = \min(r_1,r_2,r)$. Note that this prescription has also been put forward as the natural scale for the BK equation at next-to-leading order~\cite{Balitsky:2008zza}. The variable-number-of-flavours scheme is used with the same parameters as in our previous work~\cite{Cepila:2018faq,Bendova:2019psy}. 

\subsection{Glauber-Gribov approach to the nuclear dipole amplitude}
Following~\cite{Armesto:2002ny}, one can use the solution of the BK equation for the case of a proton target to obtain the dipole scattering amplitude for a nuclear target by using a Glabuer-Gribov approach
\begin{equation}
N^A(r,b,Y) = \Big[ 1-\exp\Big(-\frac{1}{2}T_A(b)\sigma_{q\bar q}(Y,r)\Big) \Big],
\end{equation}
with
\begin{equation}
\sigma_{q\bar q}(Y,r)=\int{\rm d}^2\vec{b} 2N^p(r,b,Y).
\end{equation}
This approach has been used in other studies, e.g.  those reported in~\cite{Armesto:2002ny,Cazaroto:2008iy} (see also~\cite{Marquet:2017bga} for a more general approach that reduces to the Glabuer-Gribov  case for large nuclei).
The nuclear thickness function $T_A(b)$ is obtained from a Woods-Saxon distribution for the nuclear matter density 
\begin{equation}
\rho_A(x,y,z) = \rho_0\frac{1}{\exp\left[(r-{\rm R})/{\rm a}\right]+1},
\label{eq:WS}
\end{equation}
 (where $r\equiv\sqrt{x^2+y^2+z^2}$) by integrating it over the longitudinal coordinate $z$
 \begin{equation}
 T_A(b) = \int\limits_{-\infty}^{+\infty} dz \rho_A(x,y,z),
 \end{equation}
with the $x$ and $y$ coordinates in the impact-parameter plane. It is normalised according to $\int{\rm d}^2\vec{b}\;T_A(b) = A$. (See for example~\cite{Loizides:2017ack} for full details on the formalism.) The values of the Woods-Saxon parameters are given in Table~\ref{tab:para}.  This approach is denoted as b-BK-GG in what follows.

\begin{table}[t!]
\caption{
Values of the parameters of the Wood-Saxon distribution, see Eq.~(\ref{eq:WS}), used in the computations reported in this text, and the value of the $Q^{2}_{s0}(A)$ parameter obtained as explained in the text. The Wood-Saxon parameters are taken from~\cite{DeJager:1987qc}.
\label{tab:para}}
\begin{tabular}{lrrrrr}
 \hline
 Nucleus & \ \ \ $A$&\ \ \ R (fm) & \ \ \ a (fm)  & \ \ \ 
 $\rho_0$ (fm$^{-3}$)  & \ \ \ $Q^{2}_{s0}(A)$ (GeV$^2$) \\
 \hline
Al & 27  & 2.84 & 0.569 & 0.2015& 0.315 \\
Ca & 40 & 3.51 & 0.563 & 0.17611 & 0.341 \\
Fe & 56  & 3.980  & 0.569 & 0.17655 & 0.390\\
Cu & 64  & 4.2 & 0.569 & 0.1746 & 0.409  \\
W  & 184  & 6.510 & 0.535 & 0.1493& 0.553 \\
Pb & 208  & 6.624 & 0.549 & 0.16& 0.609 \\
\hline
\end{tabular}
\end{table}

\subsection{Initial conditions for the nuclear targets}
To solve the BK equation an initial condition is needed. In our previous work~\cite{Cepila:2018faq,Bendova:2019psy} we introduced a new functional form for the initial condition given by 
\begin{equation}
\label{initialeq}
N^p(r, b,Y=0) =  1 - \exp\left(-\frac{1}{2}\frac{Q^2_{s0}}{4}r^2 T_p(b_{q_1},b_{q_2})\right),
\end{equation}
where $Q^2_{s0}$ is a free parameter representing the saturation scale at a zero impact parameter, and $b_{q_i}$ are the impact parameters of the quark and anti-quark forming the dipole.

For the case of the proton, we assumed a Gaussian like distribution which leads to
\begin{equation}
T_p(b_{q_1},b_{q_2})= \left[\exp\left(-\frac{b_{q_1}^2}{2B_G}\right) + \exp\left(-\frac{b_{q_2}^2}{2B_G}\right)\right].
\end{equation}
The parameter $B_G$ was set to 3.2258 GeV$^{-2}$, while $Q^2_{s0}$ took the value 0.496 GeV$^2$. With these values a satisfactory description of HERA and LHC data on the proton structure functions and exclusive production of vector mesons is achieved~\cite{Cepila:2018faq,Bendova:2019psy}.

\begin{figure}[t!]
\centering
 \includegraphics[width=0.48\textwidth]{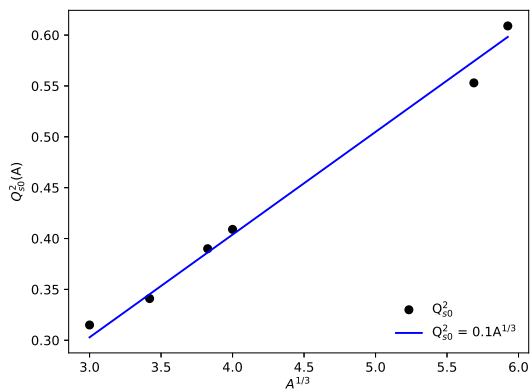}
 \caption{Dependence of the $Q^{2}_{s0}(A)$ parameter as a function of $A^{1/3}$ (solid bullets) compared with a linear function (blue line). See text for details.
  \label{fig:qs0a_vs_a}}
\end{figure}

Here, we follow a similar approach for the nuclear case, but assuming a Woods-Saxon instead of a Gaussian distribution:
\begin{equation}
\label{eq:nuclear_ic}
N^A(r, b,Y=0) =  1 - \exp\left(-\frac{1}{2}\frac{Q^2_{s0}(A)}{4}r^2 T_A(b_{q_1},b_{q_2})\right),
\end{equation}
with
\begin{equation}
T_A(b_{q_1},b_{q_2})= k \left[T_A(b_{q_1})+T_A(b_{q_2})\right].
\end{equation}
where the factor $k$ ensures that $kT_A(0)=1$. This approach is denoted as b-BK-A in what follows. 

As the nuclear parameters are already fixed, the only free parameter is $Q^2_{s0}(A)$. We have fixed  these parameters using $N(r,b,Y=0)$  where $Y=\ln(x_0/x)$ with  $x_0\equiv0.008$. This dipole scattering amplitude at the initial rapidity is used to compute structure functions and to compare them  with the predictions obtained using the EPPS16 nuclear parton distributions~\cite{Eskola:2016oht}. 

In detail, we have varied the value of the $Q^2_{s0}(A)$ parameter in order to get a small  relative deviation from the  structure function $F_2(x = 0.008, Q^2)$ as predicted by the EPPS16 PDFs. The comparison is done for the following values of the photon virtuality: $Q^2 \in$ [3.5, 4.5, 6.5, 8.5, 10, 12, 15, 18, 22, 27] GeV$^{2}$ to avoid the nonperturbative region at very low $Q^2$ and to stay in the region of virtualities where the BK equation is expected to work the best. We have used  LHAPDFs~\cite{Buckley:2014ana} to obtain the  PDF sets and the APFEL software~\cite{Bertone:2013vaa, Carrazza:2014gfa} for the computation of the structure function. The values obtained for $Q^{2}_{s0}(A)$ by this procedure are reported in the last column of Table~\ref{tab:para}.

Interestingly,  this parameter follows a linear behaviour as a function of $A^{1/3}$ as shown in Fig.~\ref{fig:qs0a_vs_a}. This opens the possibility of studying other nuclei for which there is currently no information in the EPPS16 set of parton distributions.

\section{Behaviour of the dipole scattering amplitude
\label{sec:DipoleAmplitude}}
\begin{figure}[t!]
\centering
 \includegraphics[width=0.48\textwidth]{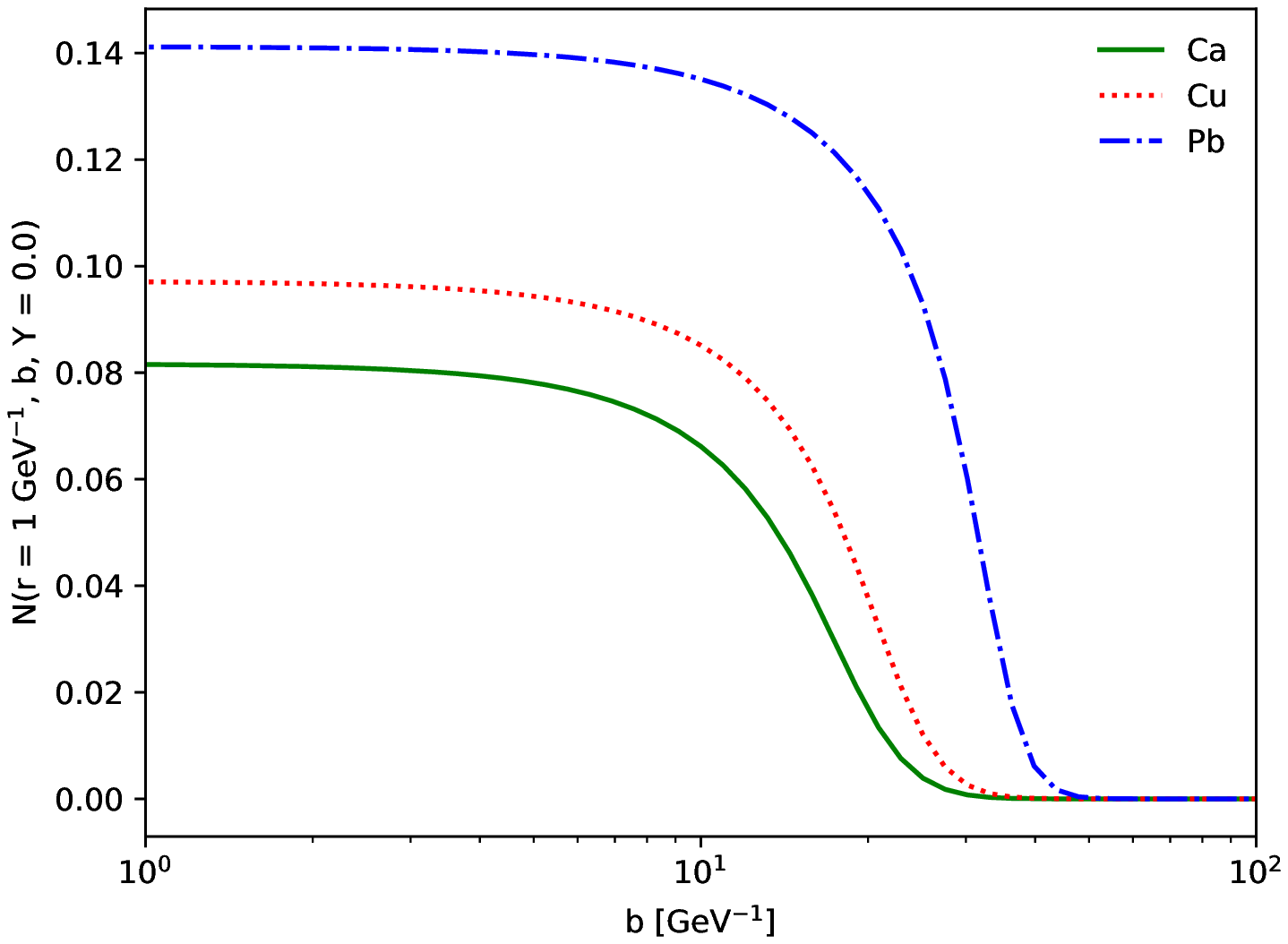}
  \includegraphics[width=0.48\textwidth]{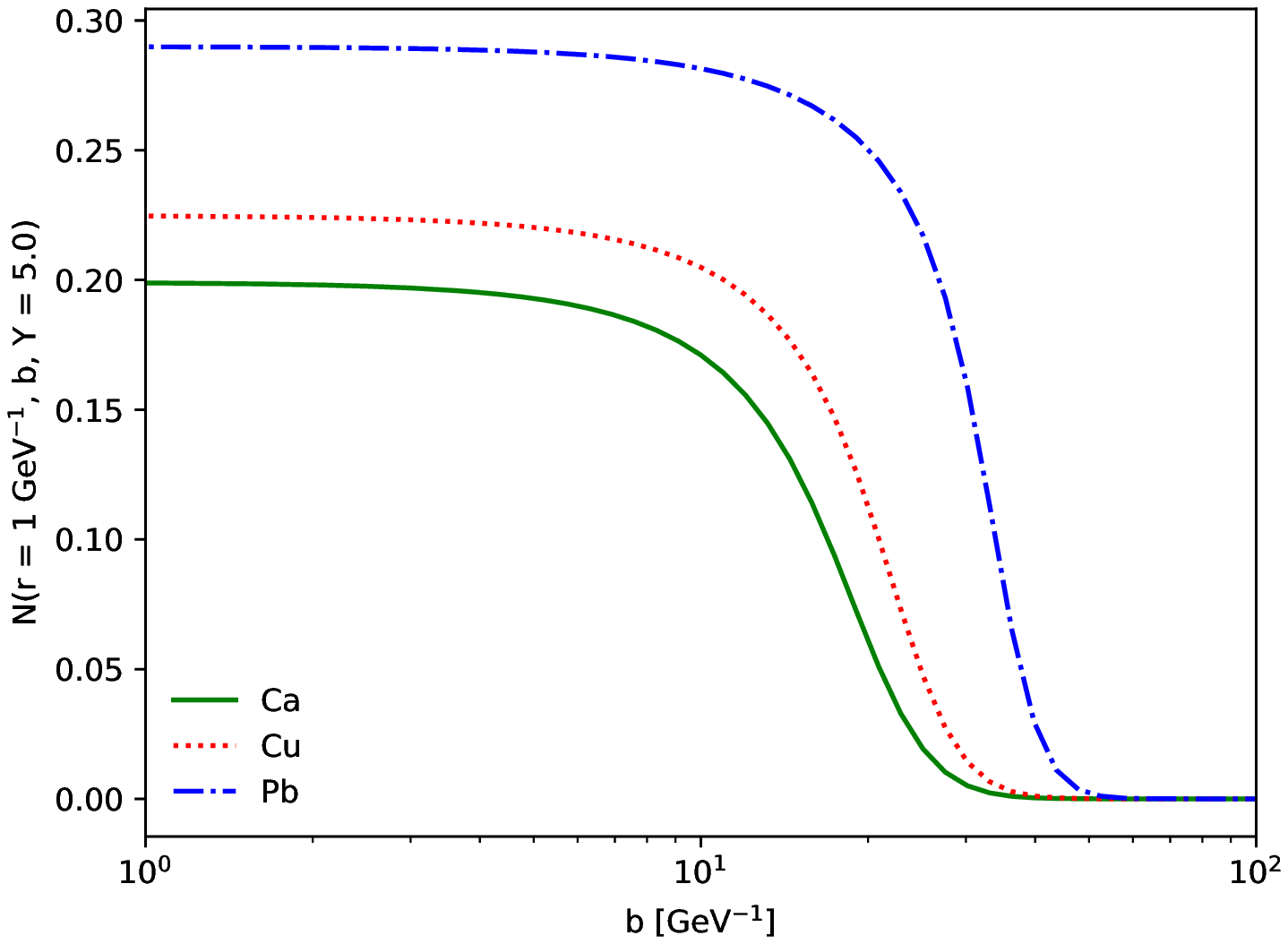}\\
 \includegraphics[width=0.48\textwidth]{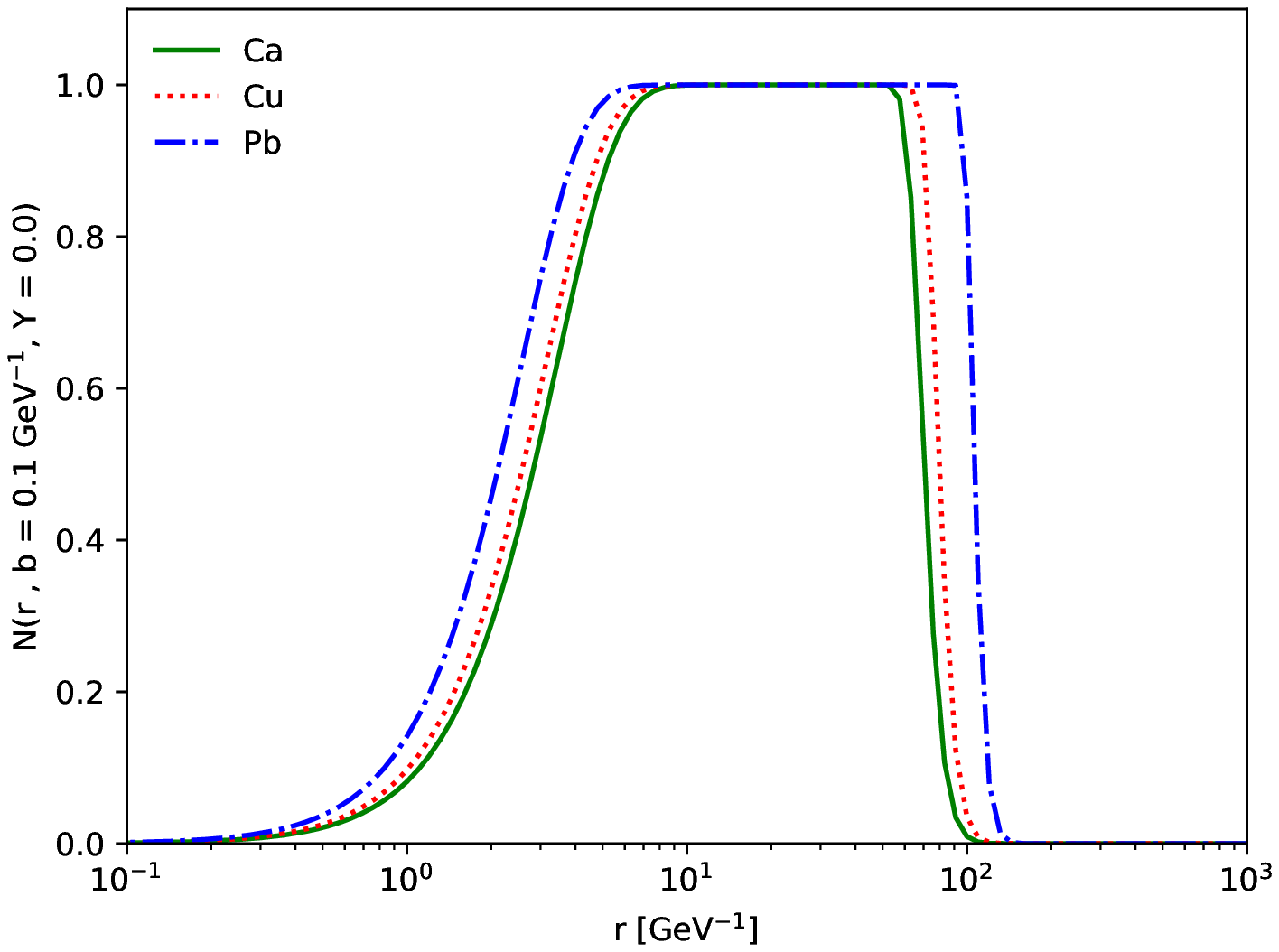}
  \includegraphics[width=0.48\textwidth]{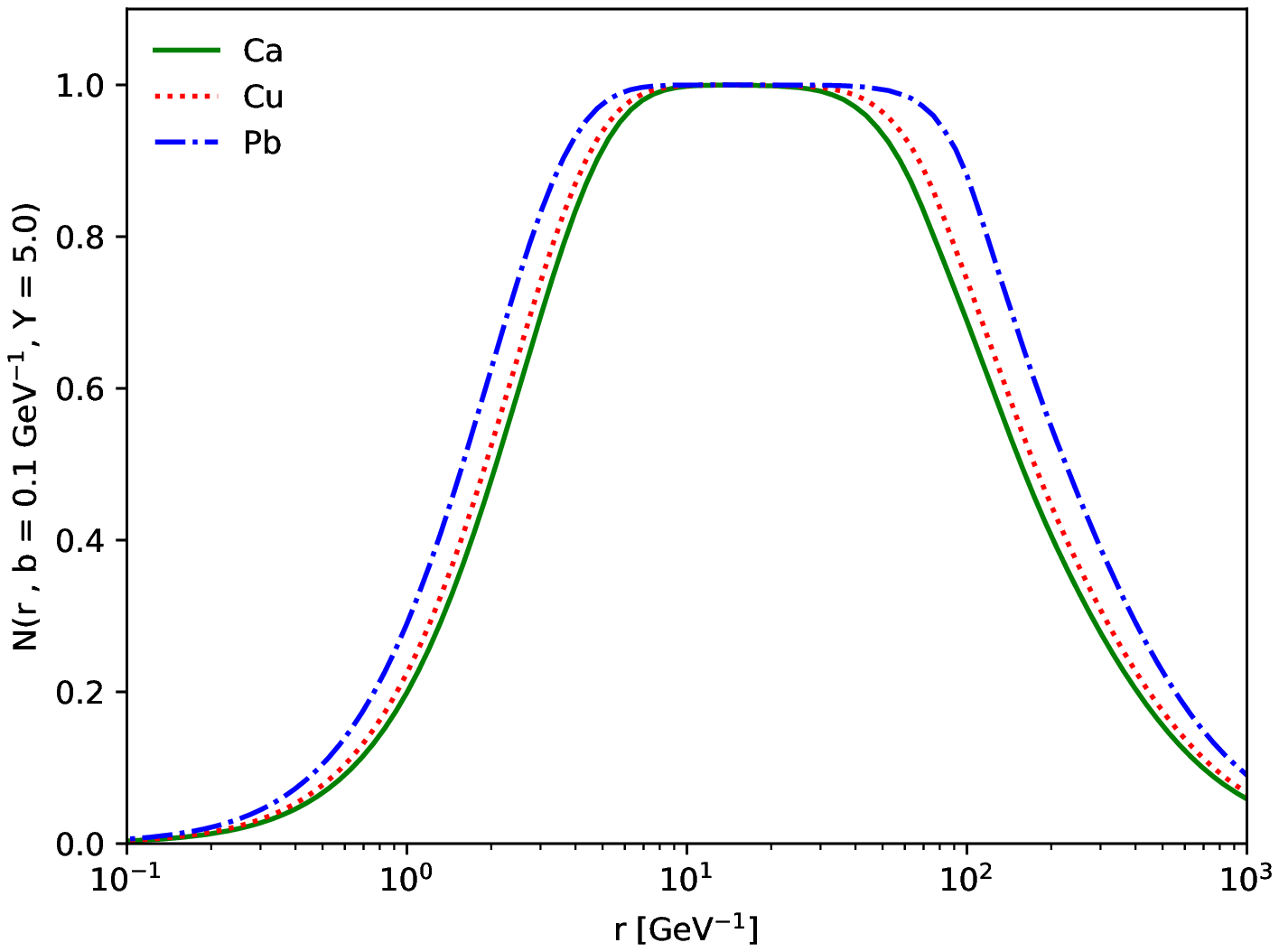}
 \caption{Dipole scattering amplitude in the b-BK-A approach at the initial condition (left) and after evolution to $Y=5$ (right) as a function of the impact parameter for a dipole size $r=1$/GeV (upper panels), and as a function of the dipole size for an impact parameter $b=0.1$/GeV (lower panels).
  \label{fig:nucelar_amplitude}}
\end{figure}

The dipole scattering amplitude in the b-BK-A approach computed using the colinearlly improved kernel with the initial condition given by Eq.~(\ref{eq:nuclear_ic}) is shown in Fig.~\ref{fig:nucelar_amplitude} for two values of the rapidity  and at a fixed value of $r$, respectively $b$, for three different nuclei. The chosen rapidity values are the initial condition ($Y=0$) and  $Y=5$, which corresponds to $x\approx 5.4\cdot 10^{-5}$ representing the case of a dipole scattering amplitude evolved to a rapidity of potential interest of future EICs. The impact parameter dependence is clearly different for the three depicted nuclei, reflecting their different sizes, while the shape of the amplitude as a function of $r$ is similar for the three cases. The main effects of the evolution are the growth of the profile in impact parameter, the softening of the large $r$ behaviour, and a small advancement of the wave front towards smaller dipoles. 

\begin{figure}[t!]
\centering
 \includegraphics[width=0.48\textwidth]{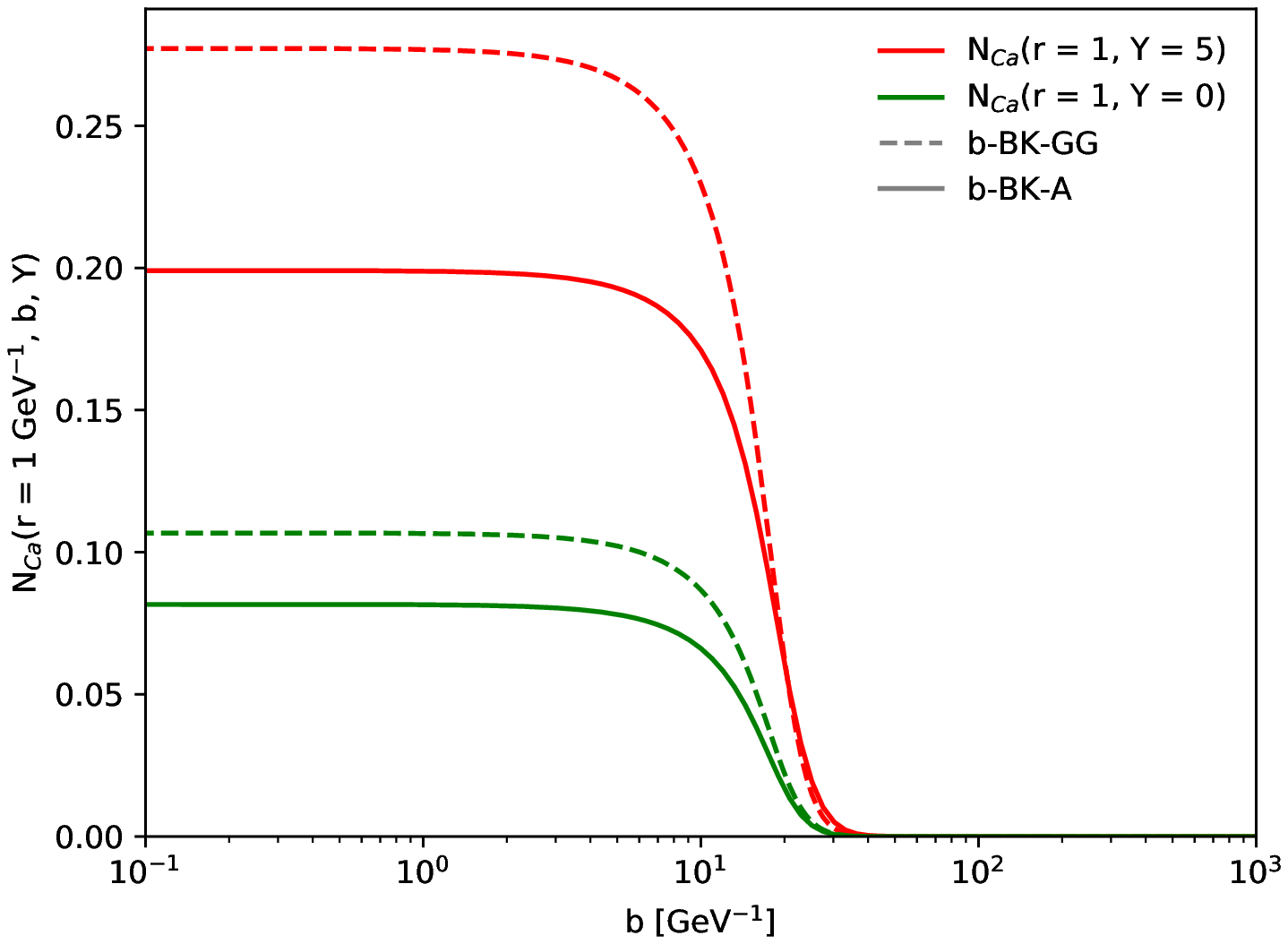}
  \includegraphics[width=0.48\textwidth]{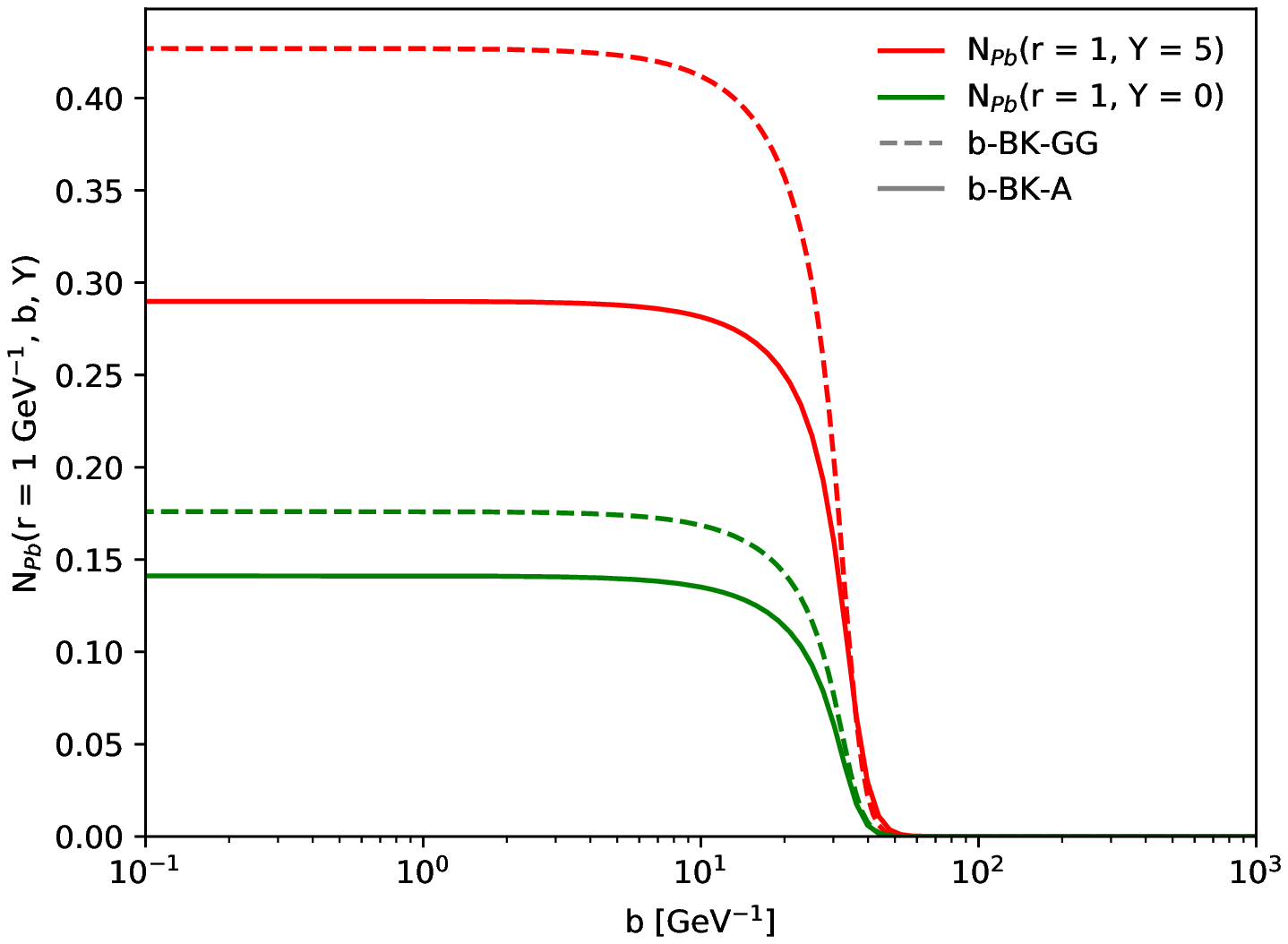}\\
 \includegraphics[width=0.48\textwidth]{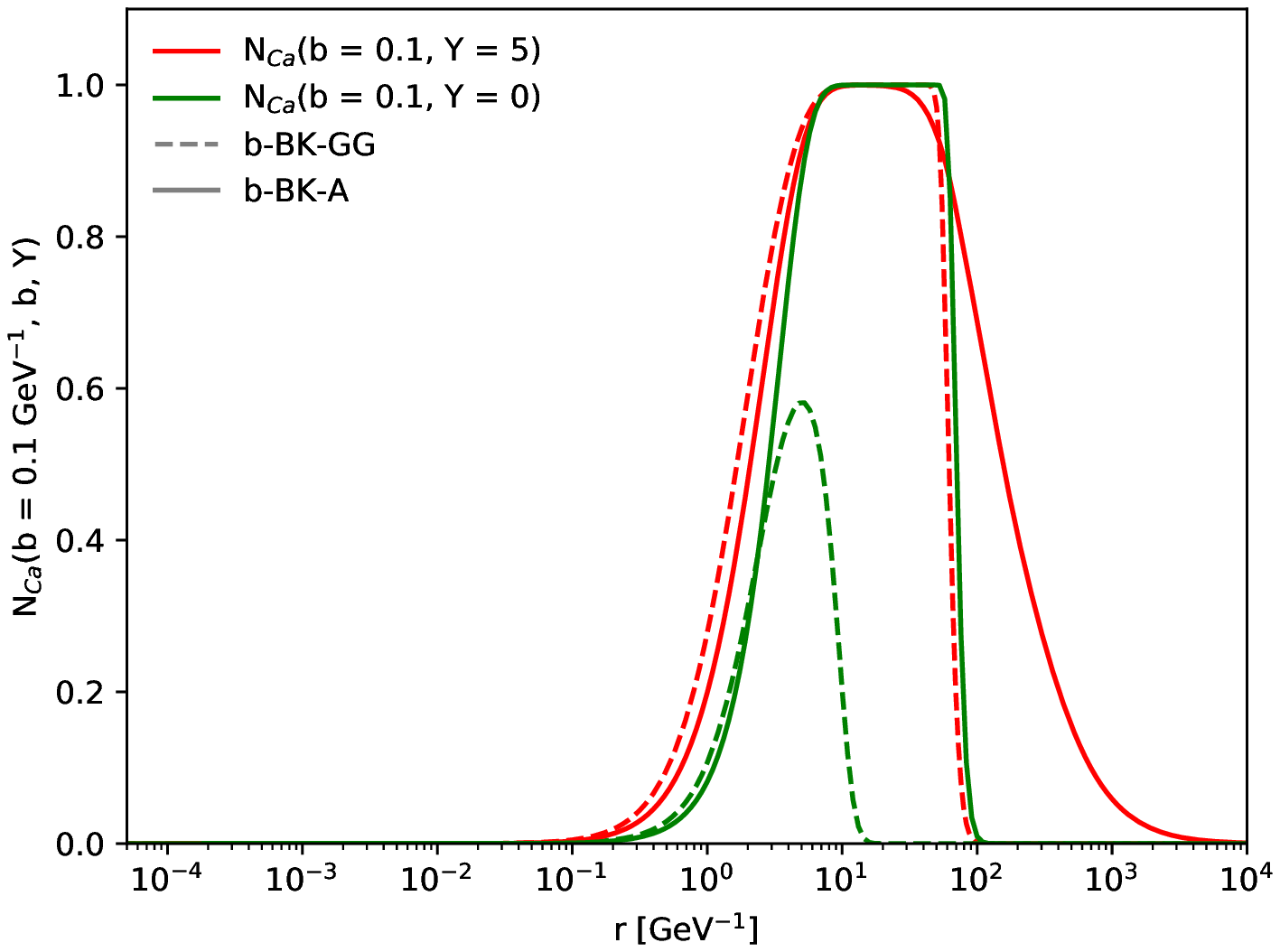}
  \includegraphics[width=0.48\textwidth]{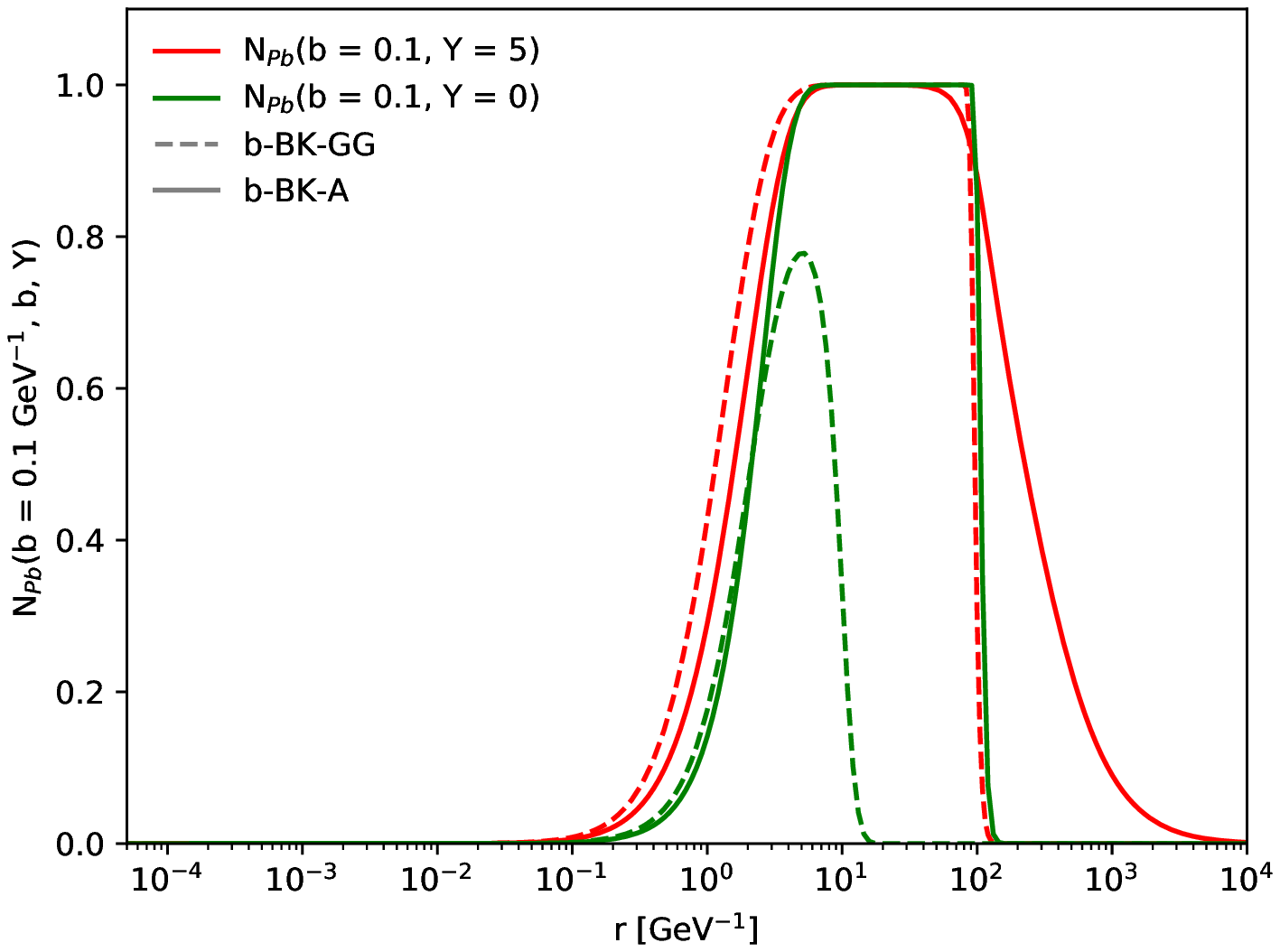}
 \caption{Comparison of the dipole scattering amplitudes computed for b-BK-A  (solid line) with the b-BK-GG approach (dashed line). The comparisons are done at $Y=0$ and $Y=5$ for two nuclei, Ca (left) and Pb (right) as a function of the impact parameter for a dipole size $r=1$/GeV (upper panels), and as a function of the dipole size for an impact parameter $b=0.1$/GeV (lower panels).
  \label{fig:nucelar_vs_galuber}}
\end{figure}

Figure~\ref{fig:nucelar_vs_galuber} shows a comparison of the two methods, b-BK-A and b-BK-GG, to compute the nuclear dipole scattering amplitude presented above.  The differences are remarkable. The absolute value of the amplitude for dipoles of size 1/GeV is substantially smaller for b-BK-A. Regarding the dependence on the dipole size at an impact parameter of 0.1/GeV the b-BK-GG approach samples dipole sizes around one order of magnitude smaller than those sampled in the b-BK-A case for large dipoles.   

\begin{figure}[t!]
\centering
 \includegraphics[width=0.48\textwidth]{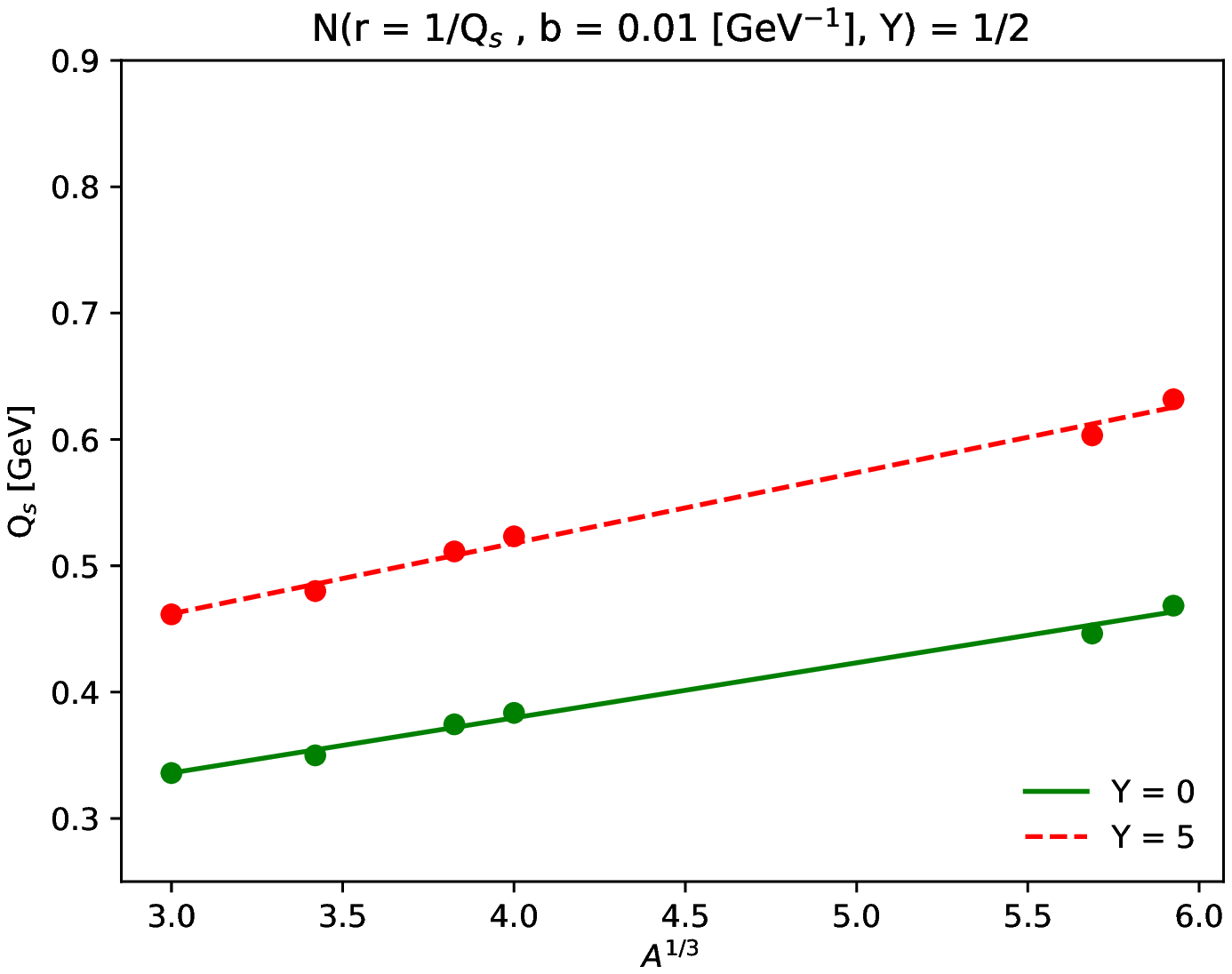}
  \includegraphics[width=0.48\textwidth]{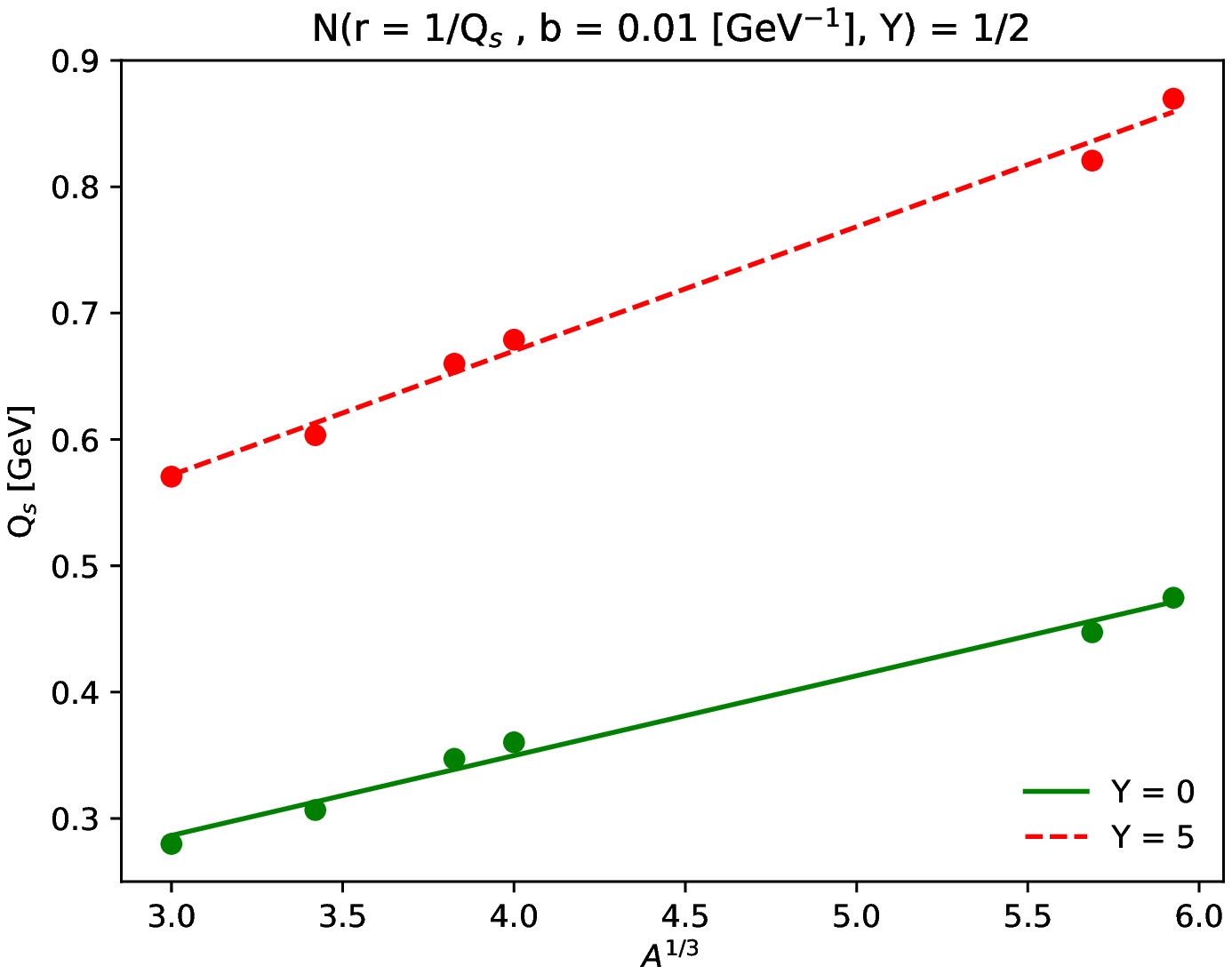}
  \caption{Saturation scale at two different rapidities for an impact parameter of 0.01/GeV for the b-BK-A (left) and b-BK-GG (right) approaches. The solid bullets are the results from the evolution and are well described by a linear function. See text for details.
  \label{fig:q2s_vs_A}}
\end{figure}

These differences between the dipole scattering amplitudes in the two approaches reflect themselves in one of the most important parameters that can be obtained from  these objects: the saturation scale and its  evolution.  As it is standard, we define the saturation scale at a given rapidity and a fixed impact parameter as the dipole size that produces a scattering amplitude equal to a constant that commonly is chosen to be one half.  Figure~\ref{fig:q2s_vs_A} shows the behaviour of the saturation scale at an impact parameter of 0.01/GeV for two rapidities as a function of $A^{1/3}$. The saturation scale shows a linear behaviour in this representation. The intercept is larger for b-BK-A with respect to b-BK-GG, while the slope is smaller. The evolution of both the intercept and the slope seems to be different in both cases. For all values of $A$ the saturation scale at $Y=5$ is smaller for b-BK-A than  for b-BK-GG predictions. Note that the figure would look the same at other values of the impact parameter due to the flat form of the dipole scattering amplitude as shown in Fig.~\ref{fig:nucelar_vs_galuber}. Only for larger values of the impact parameter, around 4 to 5/GeV, the drop at the border of the nuclei changes the behaviour of Fig.~\ref{fig:q2s_vs_A}.

\section{Predictions for nuclear structure functions and nuclear suppression factors
\label{sec:Predictions}}
\subsection{Relation between the dipole scattering amplitude and the structure function}
Using as input the dipole scattering amplitudes, the  structure function $F^A_2(x,Q^2)$ is computed as
\begin{equation}\label{F22}
 F^A_2(x,Q^2) = \frac{Q^{2}}{4\pi^{2}\alpha_{\rm em}}\int\sum_{i} d\vec{r}d\vec{b}dz \mid\Psi_{T,L}^{i}( z, \vec{r})\mid^{2} \frac{\mathrm{d}\sigma^{q\bar{q}}(\vec{r},\tilde{x})}{\mathrm{d} \vec{b}},
\end{equation}
where, following~\cite{GolecBiernat:1998js}, $\tilde{x} = x (1 + (4m^2_{q_i})/Q^2)$ with $m_{q_{i}}$  the mass of the $i$-quark. The dipole--target cross section is related to the dipole scattering amplitude by
\begin{equation}
\frac{\mathrm{d}\sigma^{q\bar{q}}(\vec{r},x)}{\mathrm{d} \vec{b}} = 2N^A(\vec{r}, \vec{b}, x).
\label{dipole-cs}
\end{equation}
Finally, the wave function representing the probability of a virtual photon splitting into a quark-antiquark dipole, and following the notation of~\cite{GolecBiernat:1998js}, is
\begin{equation}
  \mid\Psi_{T}^{i}(z, \vec{r}, Q^{2})\mid^{2} = \frac{3\alpha_{\rm em}}{2\pi^{2}} e_{q_{i}}^{2}\Big((z^{2} + (1-z)^{2})\epsilon^{2}K^{2}_{1}(\epsilon r) + m_{q_{i}}^{2}K^{2}_{0}(\epsilon r)\Big),
\end{equation}
 and 
\begin{equation}
  \mid\Psi_{L}^{i}(z, \vec{r}, Q^{2})\mid^{2} = \frac{3\alpha_{\rm em}}{2\pi^{2}}e_{q_{i}}^{2}\Big(4Q^{2}z^{2}(1-z)^{2}K^{2}_{0}(\epsilon r)\Big)
\end{equation}
for the transverse and longitudinal polarisation of the incoming photon, respectively. The total wave function is
\begin{equation}\label{psi}
 \mid\Psi_{T,L}^{i}( z, \vec{r})\mid^{2} =\mid \Psi^{i}_{T}(z, \vec{r})\mid ^{2} +  \mid\Psi^{i}_{L}(z, \vec{r})\mid^{2}.
\end{equation}
In these equations $K_{0}$ and $K_{1}$ are the MacDonald functions,  $z$ is the fraction of the total longitudinal momentum of the photon carried by the quark, $e_{q_i}$ is the fractional charge (in units of elementary charge) of quark $i$, $\alpha_{\rm em}$ = 1/137 and $\epsilon^{2} = z(1-z)Q^{2} + m_{q_{i}}^{2}$. As in our previous work~\cite{Cepila:2018faq,Bendova:2019psy} we  set the quark masses to 100\,MeV$/c^2$ for light, 1.3\,GeV$/c^2$ for charm,  and 4.5\,GeV$/c^2$ for bottom quark. As reported for example in~\cite{Iancu:2015joa} the numerical results do not depend strongly on these choices.

\begin{figure}[t!]
\centering
 \includegraphics[width=0.48\textwidth]{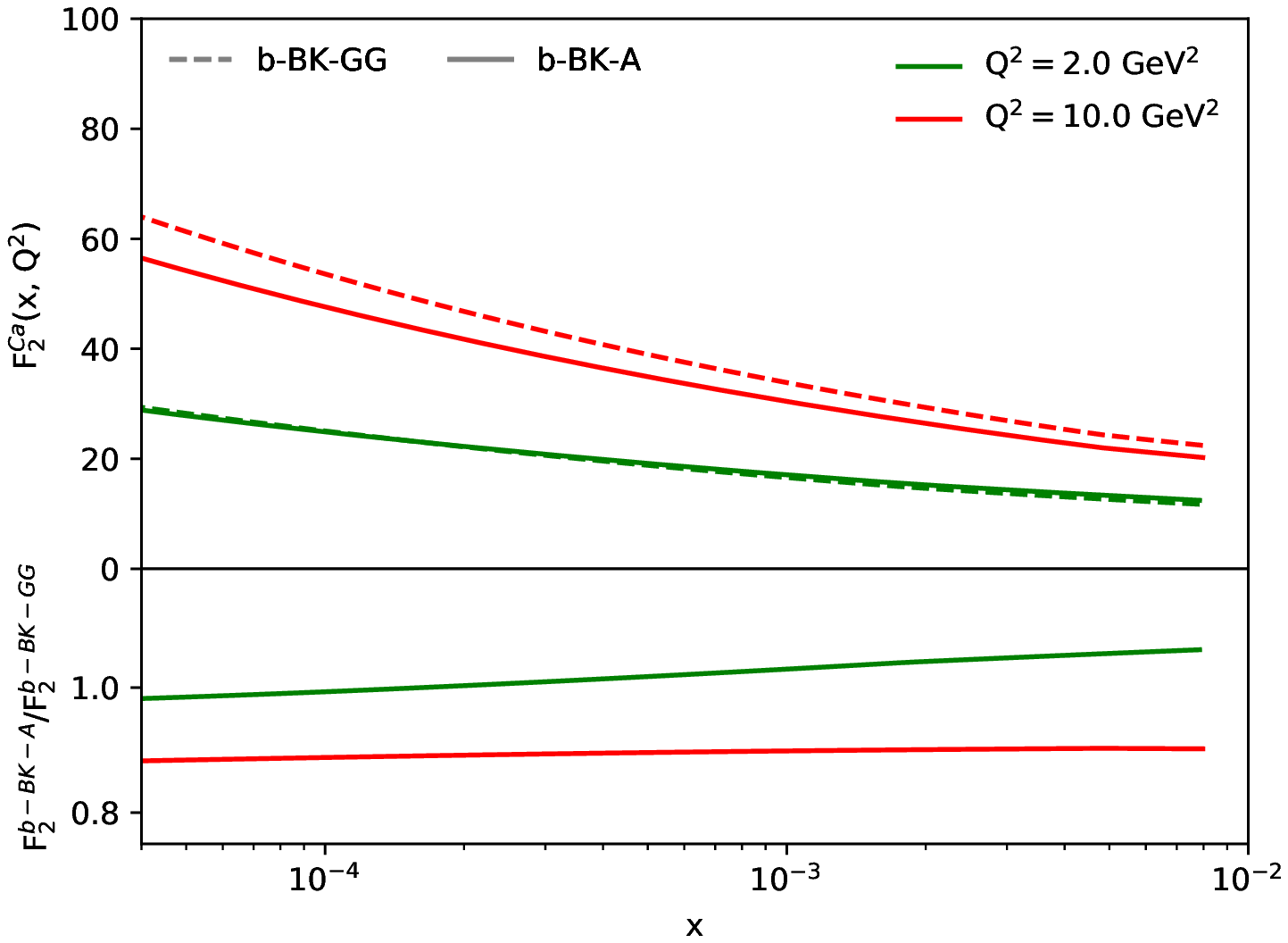}
  \includegraphics[width=0.48\textwidth]{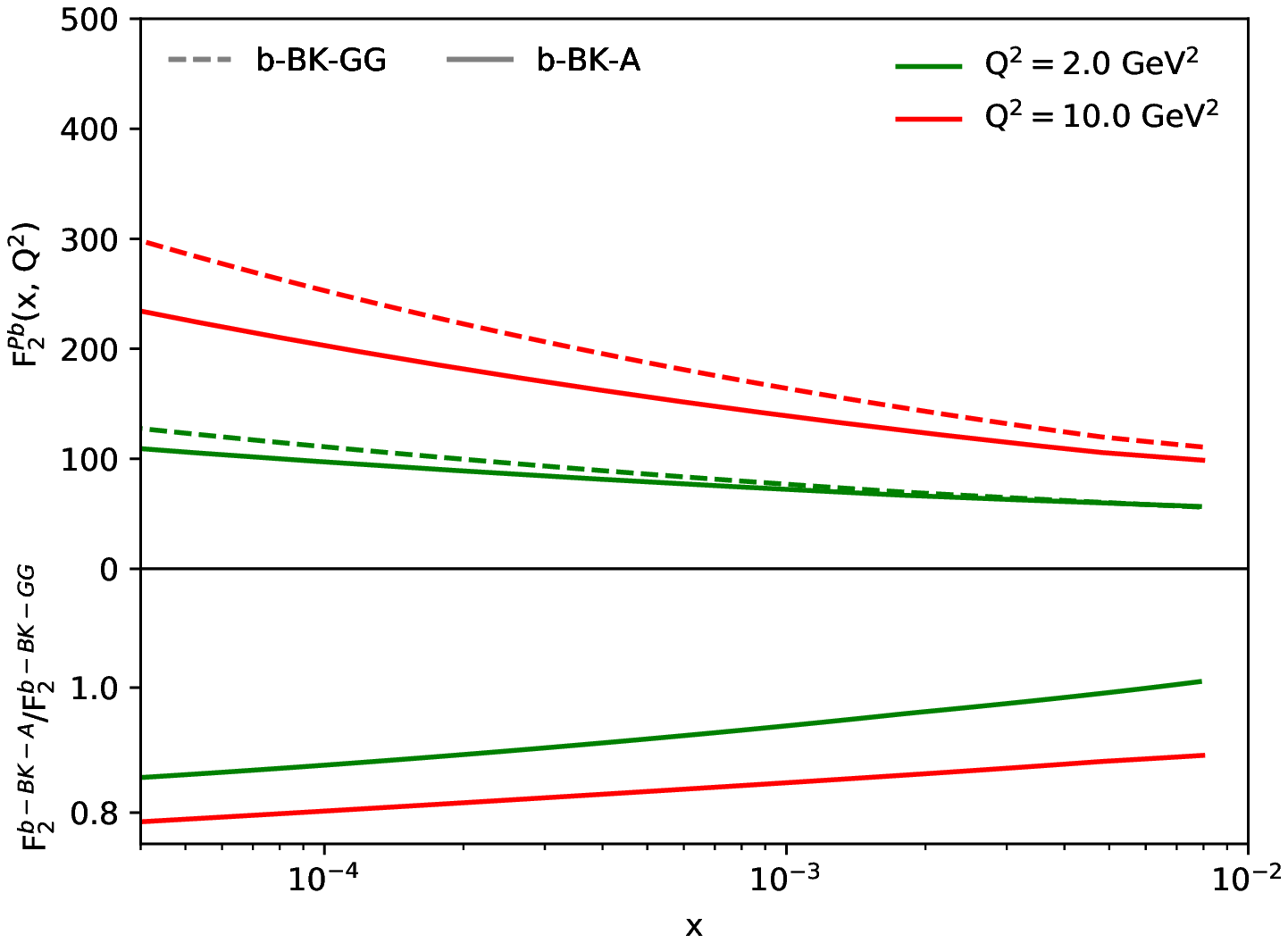}
  \includegraphics[width=0.48\textwidth]{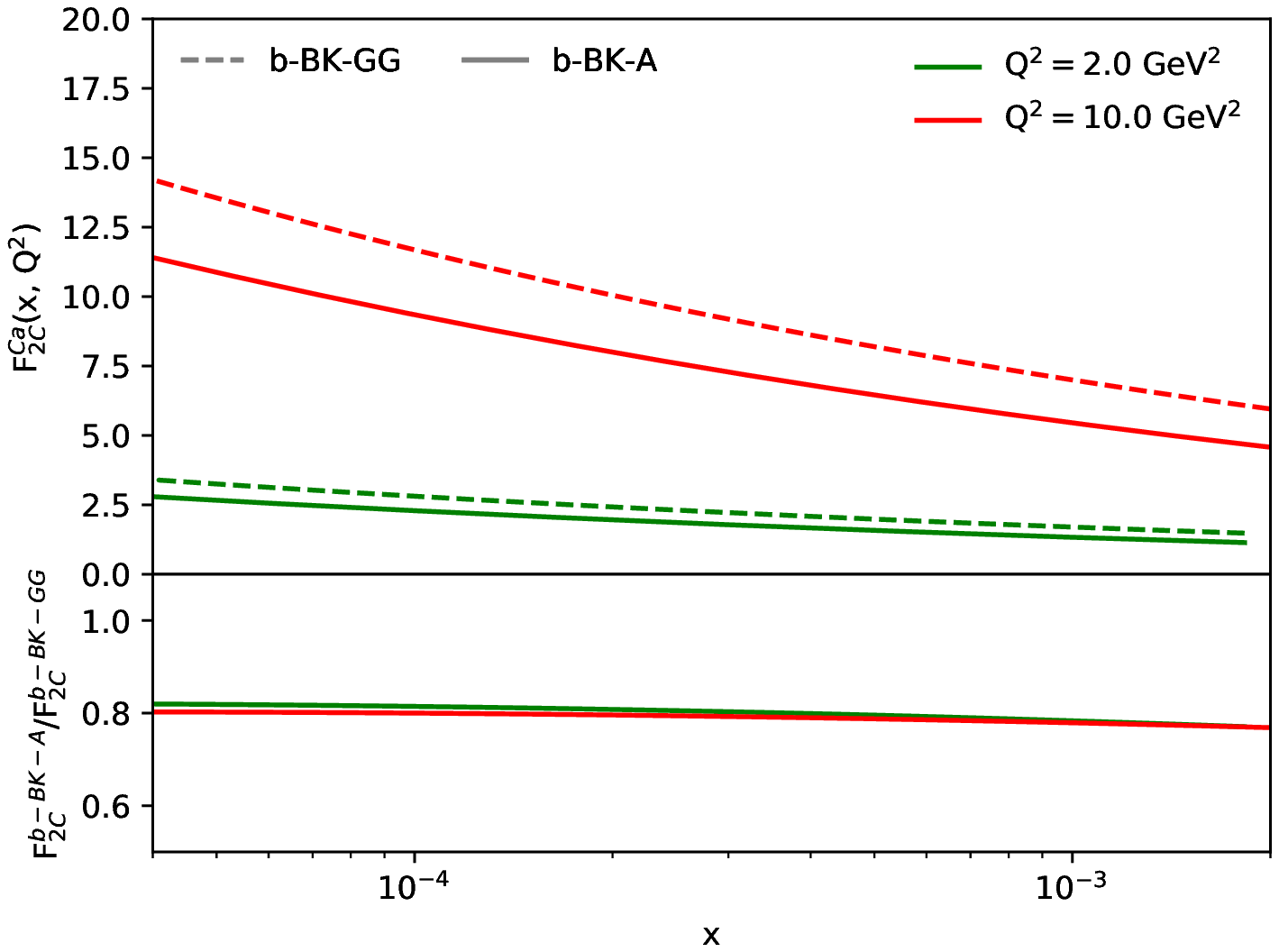}
  \includegraphics[width=0.48\textwidth]{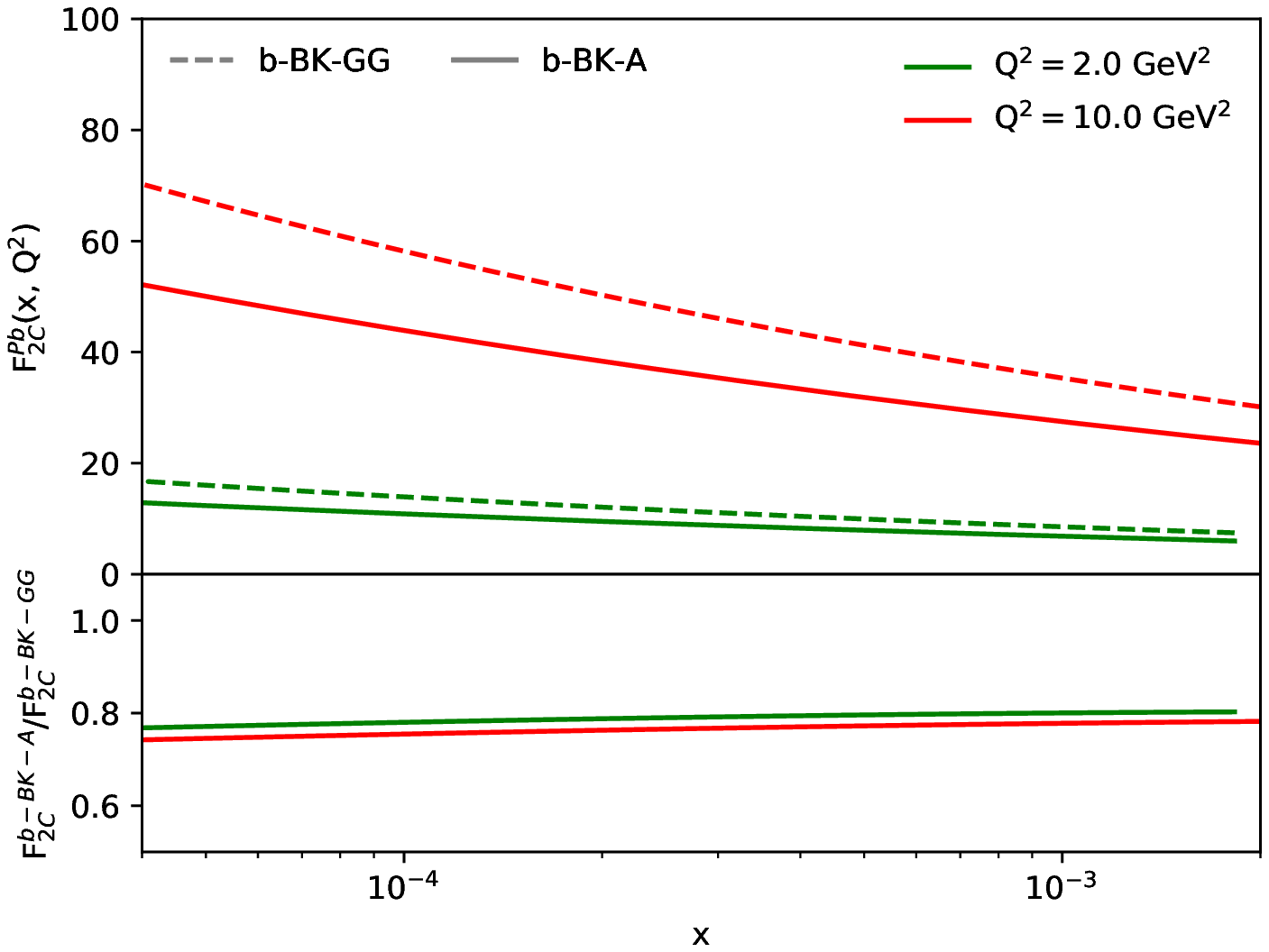}
  \caption{The upper panels show the dependence on $x$ of the nuclear structure function $F^A_2(x,Q^2)$ computed in the b-BK-A  and b-BK-GG approaches for two values of the photon virtuality $Q^2$ and two nuclei: Ca (left) and Pb (right). The
 ratio of the structure functions in the b-BK-A and b-BK-GG  approaches is shown in the lower part of those panels. The bottom panels show  the contribution of  charm, that is they show the structure function $F^A_{2C}(x,Q^2)$.
  \label{fig:F2A}}
\end{figure}

\subsection{Predictions for the nuclear structure function}
The nuclear structure functions $F^A_2(x,Q^2)$ for Ca and Pb are shown in the upper panels of Fig.~\ref{fig:F2A} as a function of $x$ for two values of the photon virtuality $Q^2$. Results in both approaches,   b-BK-A and b-BK-GG, are shown in the figure which also shows the ratio of both predicted structure functions. There is a clear difference between both sets of results. Furthermore, the difference shows a dependence on $x$, on $Q^2$, and a striking dependence on $A$, where the difference between both approaches  grows from small to large nuclei. 

A measurement of this structure function is expected to be one of the first results of any future EIC. Given the precision expected from these machines, these measurements will select which of the two approaches describes better the data. 

The lower panels of Fig.~\ref{fig:F2A} show that the difference between the b-BK-A and b-BK-GG approaches are larger for the charm structure function, $F^A_{2C}(x,Q^2)$, than for the inclusive case, and that there is a very soft dependence on kinematic variables and nucleus species. A measurement of  $F^A_{2C}(x,Q^2)$ would offer additional stringent constraints to predictions of the structure function of nuclei.

\subsection{Predictions for the nuclear suppression factor}
As a final observable we present  the nuclear suppression  factor, defined as the ratio $R_{pA}\equiv F^A_2(x,Q^2)/(A\;F^p_2(x,Q^2))$, which is expected to be unity if the structure of a free nucleon is equal to that of a bounded one. 
This ratio is the most direct way to observed nuclear shadowing, which for small $x$ is dominated by gluon shadowing and thus may be an important tool to determine the behaviour of saturation across different nuclei. 

\begin{figure}[t!]
\centering
 \includegraphics[width=0.48\textwidth]{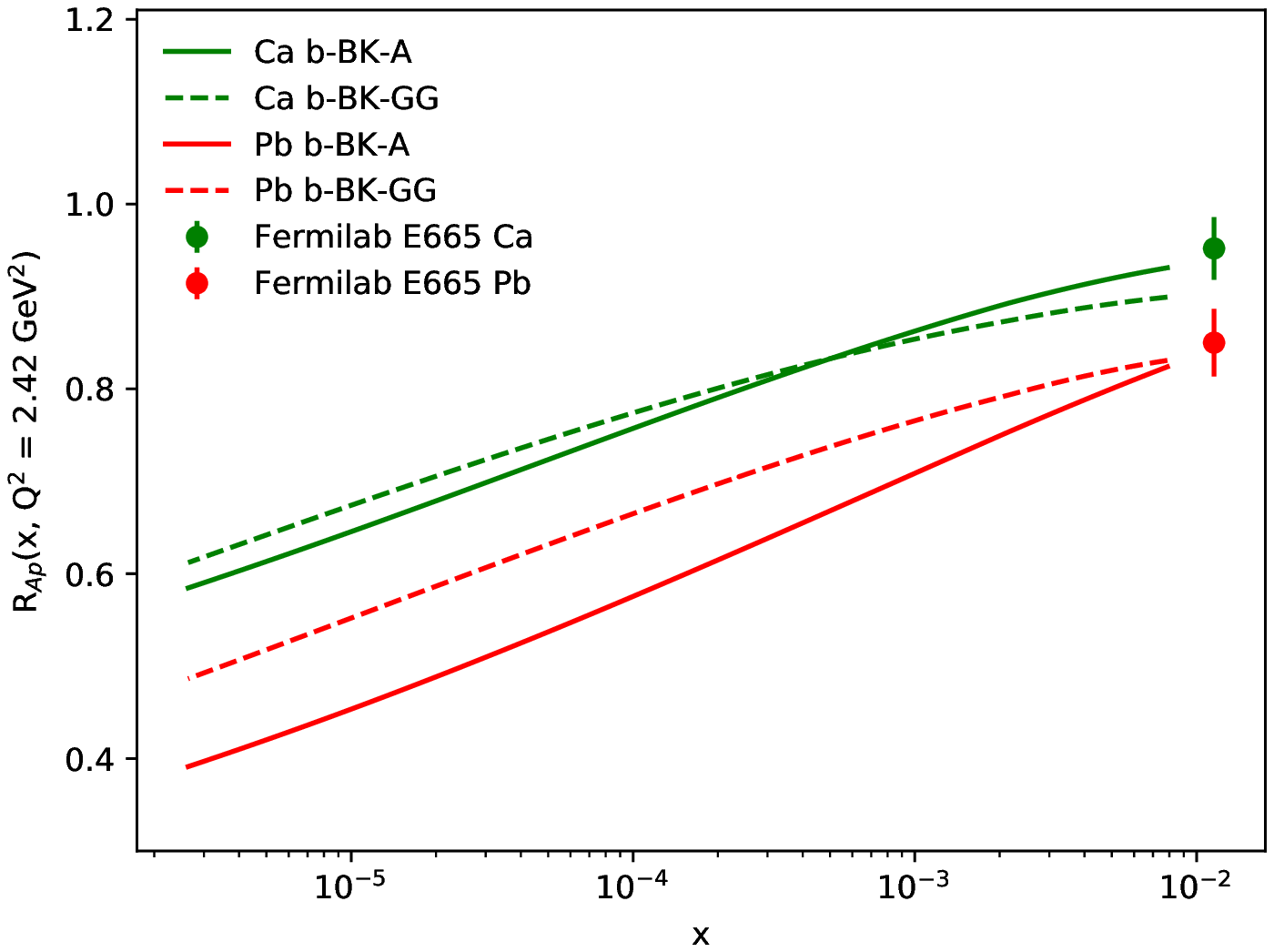}
  \includegraphics[width=0.48\textwidth]{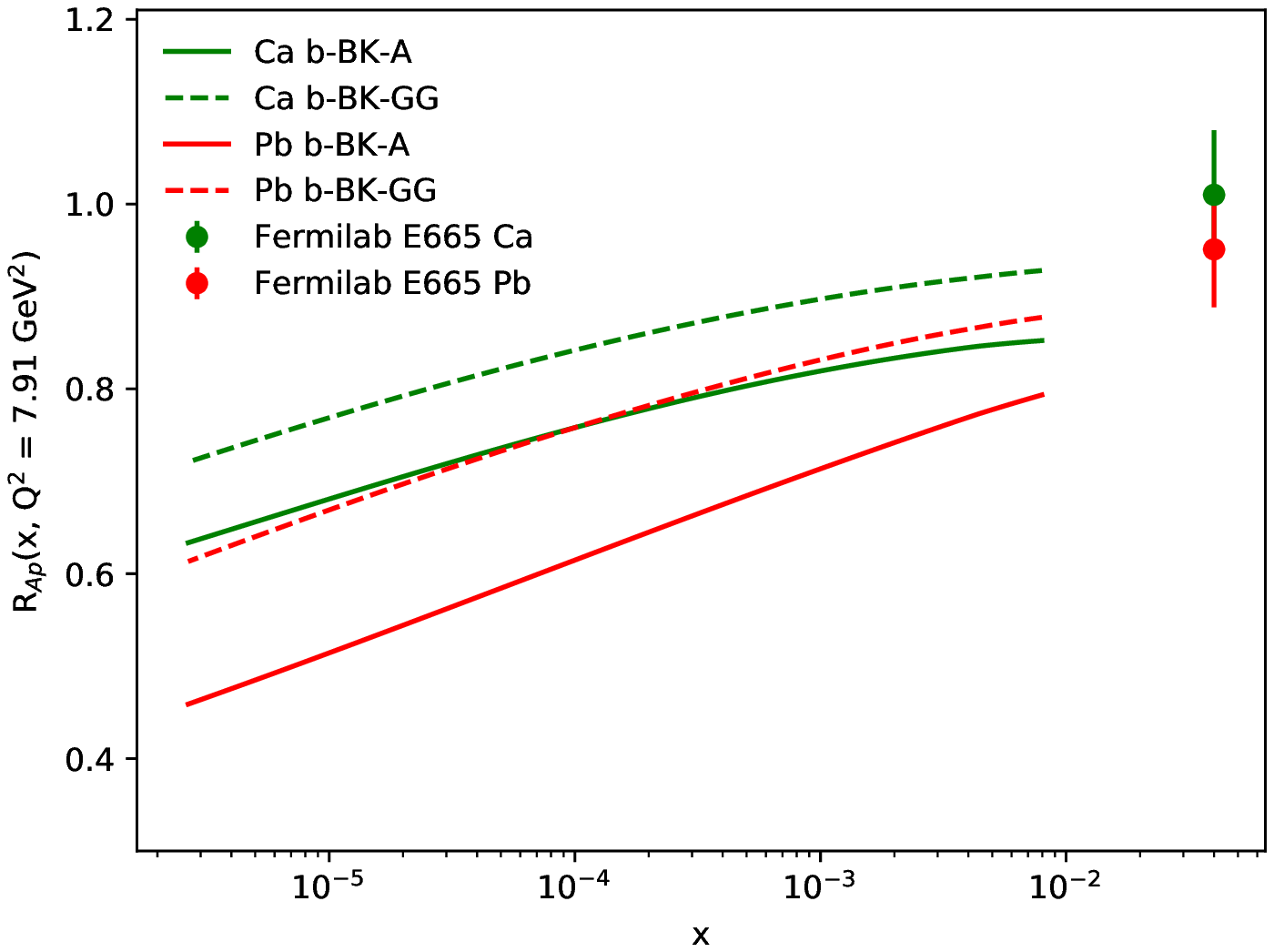} \\
 \includegraphics[width=0.48\textwidth]{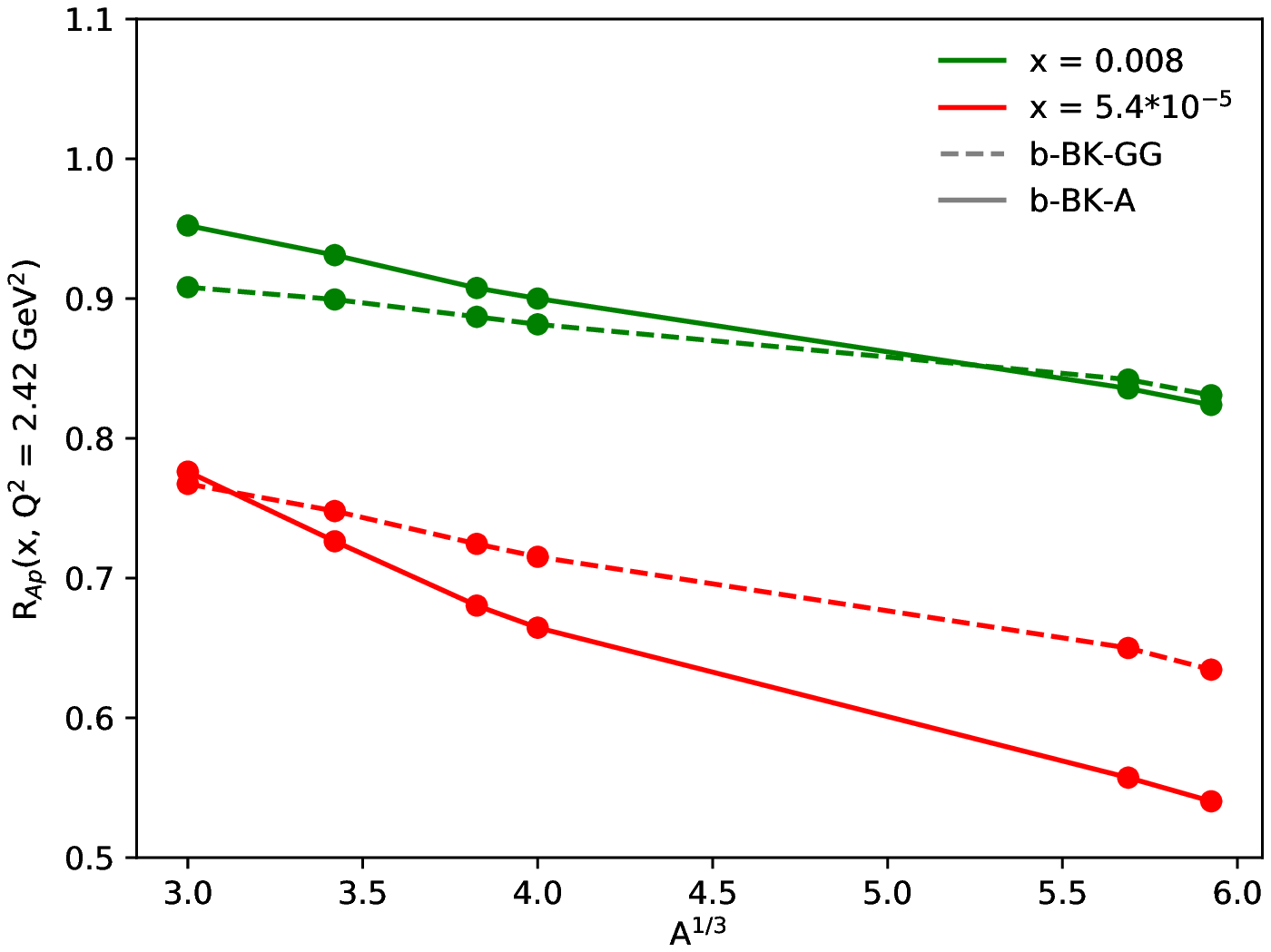}
  \includegraphics[width=0.48\textwidth]{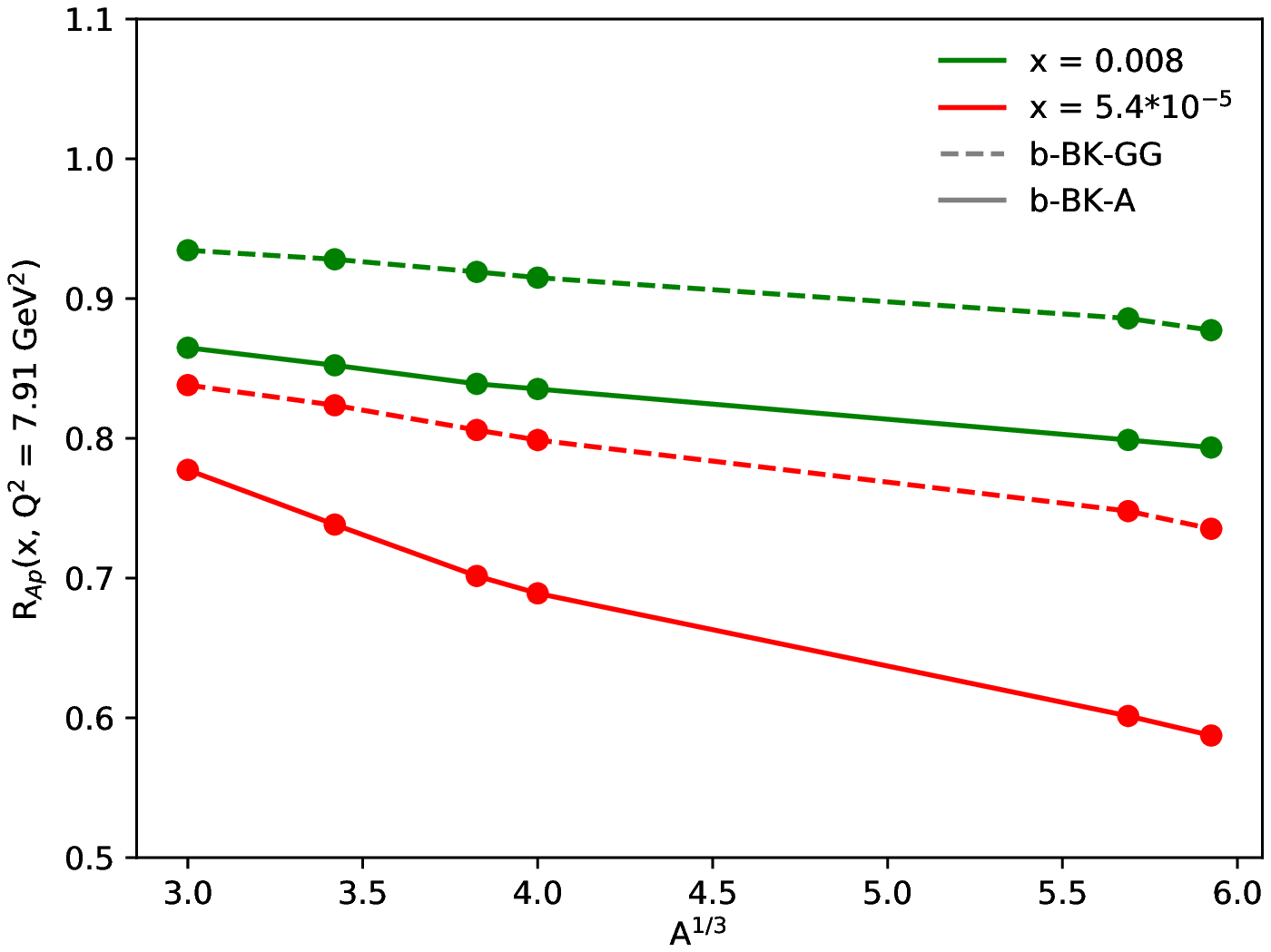}
  \caption{Nuclear suppression factor at two values of the photon virtuality $Q^2=2.42$ GeV$^2$ (left) and $Q^2=7.91$ GeV$^2$ (right) as a function of $x$ for Ca  and Pb (upper panels) and as a function of $A$ at different fixed values of $x$ (lower panels). The predictions are compared with data from~\cite{Adams:1995is}.
  \label{fig:RAp}}
\end{figure}

This factor is shown in  the upper panels of Fig.~\ref{fig:RAp} as a function of $x$. Exiting data at the same $Q^2$ from~\cite{Adams:1995is} is also shown  as a cross check of the procedure. For the $x$ dependence of $R_{pA}$ one sees a linear  decrease (in logarithmic scale) towards small $x$ for both nuclei, but the linear behaviour is reached later for the lighter nucleus, specially at higher  $Q^2$ scales.

The b-BK-A computation predicts stronger shadowing than  the b-BK-GG case with this behaviour seemingly dependent on $Q^2$.  The same figure also shows, in the bottom panels, the $A$-dependence of the nuclear suppression factor for  two $Q^2$ scales and for two values of $x$. As expected, shadowing becomes stronger as the size of the nucleus grows. The different behaviour of shadowing for different nuclei in the b-BK-A and b-BK-GG is clearly seen in this observable.

\begin{figure}[t!]
\centering
 \includegraphics[width=0.48\textwidth]{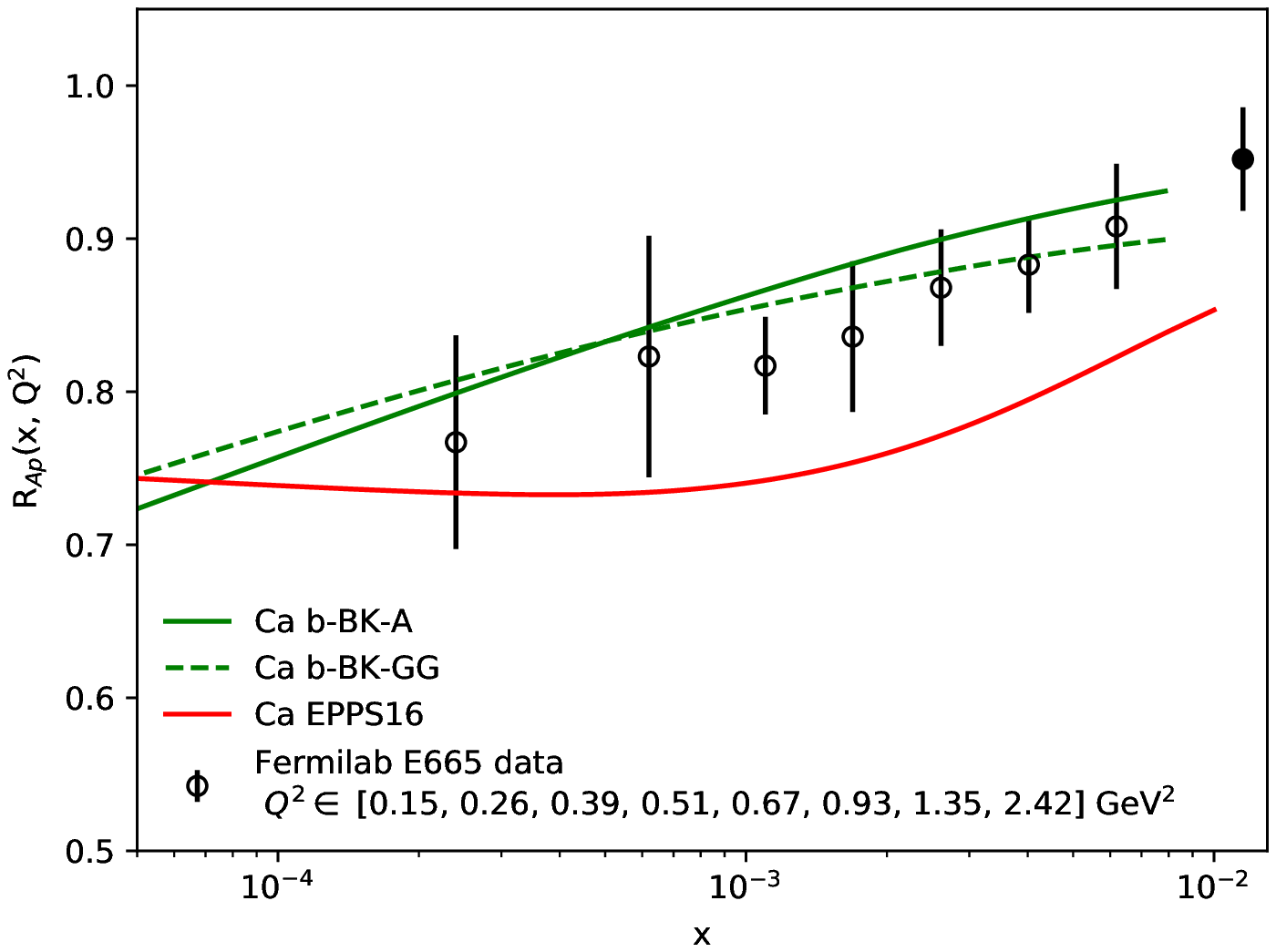}
  \includegraphics[width=0.48\textwidth]{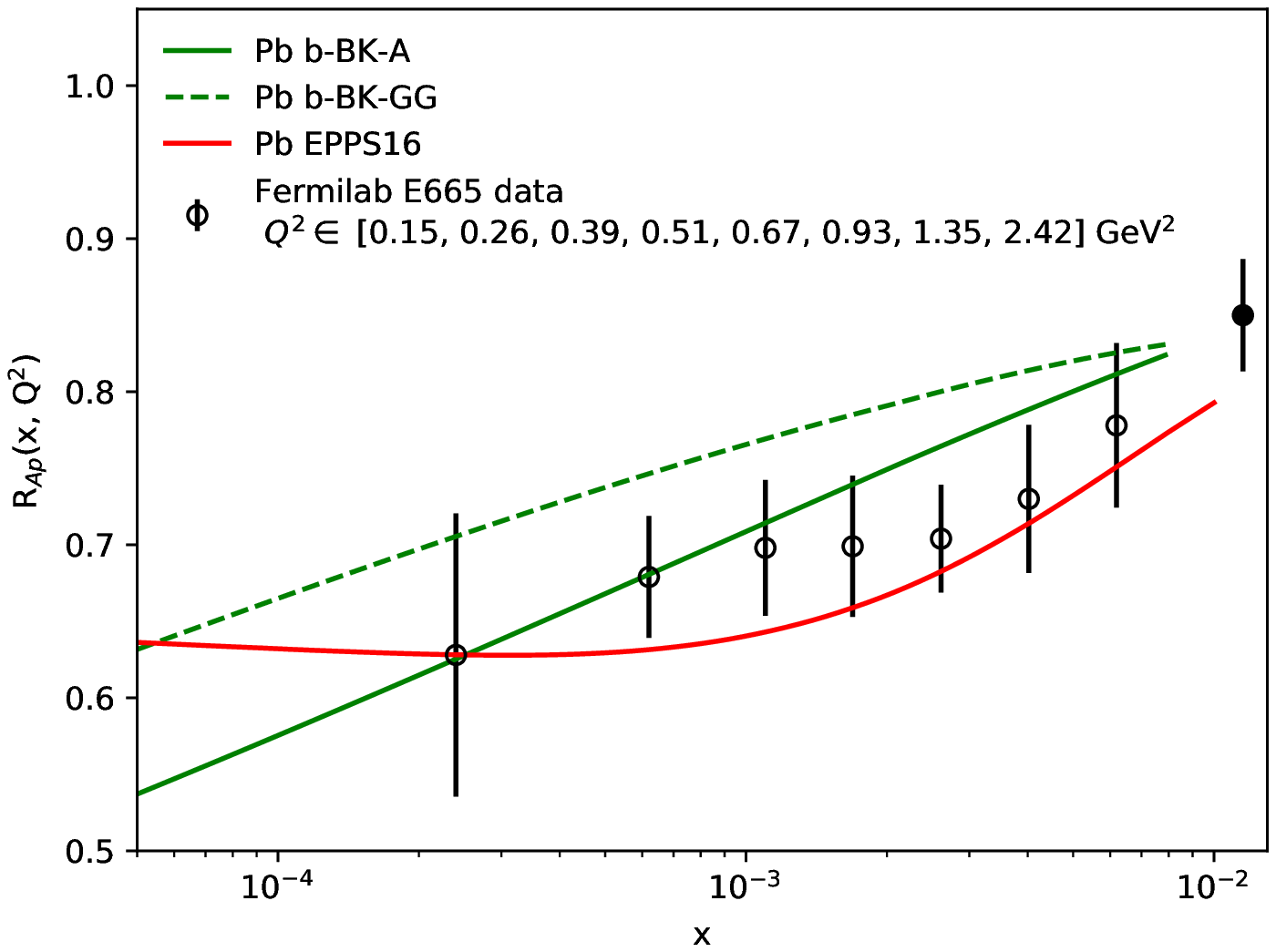} \\
  \caption{Nuclear suppression factor for Ca (left) and Pb (right) for $Q^2$ = 2.42\,GeV$^2$. The data-points correspond to $Q^2$ values of [0.15, 0.26, 0.39, 0.51, 0.67, 0.93, 1.35, 2.42] from left to right. Only the data point represented by a solid marker correspond to the $Q^2$ used for the predictions. See text for details.
  \label{fig:vsEPPS16}}
\end{figure}

Figure~\ref{fig:vsEPPS16} shows the comparison of our predictions with those obtained using EPPS16 which is considered a standard of our current knowledge of nuclear shadowing. The comparison is done for Ca and Pb as middle and large nuclei. The predictions are compared with data from~\cite{Adams:1995is}. Note that the predictions are at a $Q^2$ scale of 2.42 GeV$^2$ which we considered the lowest we would like to go to stay in a somehow perturbative scale. But the data is measured at a different $Q^2$ for each $x$ value (as illustrated by the use of empty markers for data at smaller $Q^2$). The $Q^2$ values are reported in the figure. 

Focusing on the predictions, the behaviour at small $x$ is definitely different for the EPPS16 and BK computations.  Note that the difference between EPPS16 and b-BK at the initial scale used for the BK evolution have two origins: one, that the parameter of the initial scale shown in Fig.~\ref{fig:qs0a_vs_a} is chosen by comparing with larger values of $Q^2$ than those shown in Fig.~\ref{fig:vsEPPS16}, and two that the prediction for the structure of the proton is substantially different for EPPS16 and for the b-BK approach reported in~\cite{Cepila:2018faq,Bendova:2019psy}. Given that the difference among the approaches goes beyond a normalisation factor and shows a strong $x$-dependence, data from future EICs are expected to be precise enough to decide which prediction is closer to reality.

Comparing with the currently available data, and taking into account ($i$) the different $Q^2$ in data and predictions, and ($ii$) that for measurements the values are quite low (even below what one would expect to be valid for an approach based on perturbative QCD), the b-BK-A prediction seems to do a reasonable job of describing data. The EPPS16 prediction also does quite well for  Pb, but slightly worse for Ca. The b-BK-GG prediction on the other hand is good when comparing with Ca, but it suffers a bit when compared with Pb.

\section{Summary and outlook
\label{sec:Summary}
}
The dipole scattering amplitudes, including the impact parameter dependence, for different nuclei have been obtained by solving the BK equation with the collinearly improved kernel. These amplitudes have been used to predict structure functions and nuclear saturation factors in kinematic ranges of interest for future EICs as those currently planned in the USA and at CERN. We followed two approaches: modelling the target directly as a nucleus using Wood-Saxon parameterisations (denoted as b-BK-A above), and solving for a proton and using a Glauber-Gribov prescription to go to the nuclear level (denoted as b-BK-GG above).

We find sizable differences between these approaches. These differences show a dependence on $x$, $Q^2$ and $A$ such that data from a future EIC will be able to select the most appropriate approach for the description of data. We also compared nuclear suppression factors with those predicted using the EPPS16 formalism which is taken as a standard of our current knowledge of nuclear shadowing. We find that all three approaches yield different predictions and that the b-BK-A computation seems to provide a better description of existing data.  

These studies show that the data expected from a future EIC have the capability of select the best theoretical approach and thus to advance our understanding of the nuclear structure, of shadowing, and of the high energy limit of QCD.

The dipole scattering amplitudes computed in this work  are publicly available in the website 
\url{https://hep.fjfi.cvut.cz/} along with macros and instructions to facilitate their use for anybody interested.

\section*{\label{sec:Acknowledgements}Acknowledgements}
This work was partially performed within the activities of the  Centre of Advanced Applied Sciences with the number: CZ.02.1.01/0.0/0.0/16-019/0000778. The Centre of Advanced Applied Sciences is co-financed by the European Union. This work has also been partially supported from grant LTC17038 of the INTER-EXCELLENCE program at the Ministry of Education, Youth and Sports of the Czech Republic and the COST Action CA15213 THOR. Computational resources were provided by the CESNET LM2015042 grant and the CERIT Scientific Cloud LM2015085, provided under the program Projects of Large Research, Development, and Innovations Infrastructures.

 \bibliography{F2A}

 \end{document}